# A metric Suite for Systematic Quality Assessment of Linked Open Data

Behshid Behkamal
Ferdowsi University of Mashhad, Iran
behkamal@um.ac.ir

Moshen Kahani
Ferdowsi University of Mashhad, Iran
kahani@um.ac.ir

Ebrahim Bagheri
Ryerson University, Toronto, Canada
bagheri@ryerson.ca

Majid Sazvar
Ferdowsi University of Mashhad, Iran
sazvar@alumni.um.ac.ir

*Abstract*- **The vision of the Linked Open Data (LOD) initiative is to provide a distributed model for publishing and meaningfully interlinking open data. The realization of this goal depends strongly on the quality of the data that is published as a part of the LOD. This paper focuses on the systematic quality assessment of datasets prior to publication on the LOD cloud. To this end, we identify important quality deficiencies that need to be avoided and/or resolved prior to the publication of a dataset. We then propose a set of metrics to measure these quality deficiencies in a dataset. This way, we enable the assessment and identification of undesirable quality characteristics of a dataset through our proposed metrics. This will help publishers to filter out low-quality data based on the quality assessment results, which in turn enables data consumers to make better and more informed decisions when using the open datasets.**

*Keywords:* **Metrics, RDF Datasets, Systematic Assessment, Quality Deficiencies, Linked Open Data**

## I. INTRODUCTION

Linked Open Data (LOD) provides the possibility for data providers to publicly publish their data and meaningfully link them with other data sources over the Web. The main goal of the Web of Data initiative is to create knowledge by interlinking dispersed but related data instead of linking related documents in the traditional Web.

The technology behind such interlinking of data is based on three simple principles: *i)* using URIs to name and link entities; *ii)* using the HTTP protocol for retrieval; and *iii)* using a standard model called Resource Description Framework (RDF) to describe data. This way, data can be reused and extended by other publishers and application developers. These links make the current LOD cloud, which consists of over 50 billion pieces of data represented as RDF triples covering a diverse set of domains [1]. This massive amount of data on the LOD opens up significant challenges with regards to data quality. Some of the published datasets suffer from quality problems, most of which come from the information extracted from semi-structured or even unstructured sources[1]. It is clear that such problems within the Web of Data impact the usefulness and applicability of the LOD.

Researchers have already developed several methodologies, metrics and tools to evaluate data quality in general. For example, Pipino et al [2] describe subjective and objective assessments of data quality and present three functional forms for developing objective data quality metrics including simple ratio, min/max operation and weighted average. In [3], Lee et al have proposed a methodology for the assessment of organizational

Information Quality (IQ), which consists of a systematic questionnaire to measure IQ and is accompanied by analysis techniques for interpreting the proposed IQ measures. In the area of the methodologies for data quality assessment, Batini et al [4] provide a comparative description of existing methodologies and offer a comparison of these methodologies along several dimensions, including the methodological phases and steps, the strategies and techniques, the data quality dimensions, the types of data, and, finally, the types of information systems addressed by each methodology. The database community has also developed a number of approaches such as user experience, expert judgment, sampling, parsing and cleansing techniques [5],[6] for measuring and enhancing data quality.

Despite the fact that data quality is an important requirement for the successful growth of the LOD, only a limited number of research initiatives exist, that focus on data quality assessment for the Semantic Web and specifically for the LOD. To the extent of our knowledge, there is only one work, which presents a systematic review of the approaches for assessing LOD data quality as well as a comprehensive list of dimensions and metrics [1].

Based on our experience in publishing academic open data [7],we recognize that the quality of published data has roots in the quality of the data sources from which the data has been extracted. Thus one of the better strategies to avoid quality issues is to assess the quality of a dataset before it is published. This will help publishers to filter out low-quality data based on quality assessment results, which in turn enables data consumers to make better and more informed decisions when using shared datasets.

Given the fact that Resource Description Framework (RDF) offers a standardized means of representing information on the Web of Data, and given the availability of tools for converting other data formats into RDF[8], the scope of our current work is defined to cover RDF datasets. Thus, we identify quality issues of a given RDF dataset that can be avoided before release, by proposing a set of metrics to address such problems. This way, we are able to assess the quality of datasets before publishing the data by observing the measured values of the relevant metrics. Therefore, the main research question that we will be trying to answer in this paper is the following: 'Can a set of metrics be defined to identify and quantitatively measure quality issues of an RDF dataset?' The idea behind this research question is derived from the area of software quality measurement; where metrics are defined as measurable syntactic aspects of software artifacts, e.g., Lines of Code. The novel contributions of our work can be summarized as follows:

1. We identify significant quality issues in published datasets that can be avoided prior to publication on the LOD cloud;

2. We formally define a set of automatically measureable metrics for quality assessment of a dataset in the initial phases of publication;

3. We theoretically validate the proposed metrics and evaluate their applicability by measuring the quality deficiencies of published datasets;

4. We introduce a novel approach for data quality assessment, which has its roots in measurement theory and software measurement techniques.

The rest of this paper is organized as follows: first, data quality research in the area of the LOD is reviewed in Section 2. In Section 3, our approach for proposing metrics starts by identifying significant quality issues in RDF datasets, followed by the development of suitable metrics to address the issues, which are further theoretically validated. Empirical evaluation of the developed metrics is provided in Section 4, and then discussion on the results of our observations is presented in Section 5. Finally, the paper is concluded by presenting prospects for future work in Section 6.

## II. LITERATURE REVIEW

In this section, we classify the related literature into three main groups: I) data quality assessment frameworks, ii) quality problems of published data, and iii) tools and applications for validation of RDF datasets.

In the context of quality assessment frameworks, Hartig and Zhao [9] have proposed a framework to assess the information quality of Web data sources based on provenance information. Furthermore, Bizer and Cyganiak have developed a framework, called WIQA, which filters poor information in Web-based information systems according to user defined quality requirements [10]. As mentioned before, only one work has been already published that focuses on the quality dimensions and metrics for LOD and presents a systematic review of approaches for assessing the data quality of LOD as well as a comprehensive list of dimensions and metrics [1].The other approaches of the first group have used Semantic Web technologies to identify and correct data quality issues. For instance, the approach proposed by [11] has exploited both domain and background knowledge to detect data deficiencies in metadata including spurious annotations, data inconsistencies, duplicates, ambiguous and inaccurate data. A more recent work has used ontologies to annotate incorrect data, such as redundant instances or incorrect attribute value combinations to train detection algorithms for automated identification of data quality problems in cancer registries and data sources from the energy industry [12].

The second group of related work investigates quality problems in published datasets. The most comprehensive work in this group is conducted by the Pedantic Web Group [13]that classifies quality problems of the published linked datasets and discusses common errors in RDF publishing, their consequences for applications, along with possible publisher-oriented approaches to improve the quality of machine-readable and open data on the Web. In another work, Furber and Hepp propose an approach to evaluate the quality of datasets using SPARQL queries in order to identify quality problems such as missing literal values or data type properties, illegal literal values, and functional de-pendency violations. Using this approach, the authors identify quality problems of already available datasets such as Geonames and DBPedia [14].

The last group of work includes some tools for validating RDF datasets, each with its own error-checking functionalities. Some, which are available online, accept an RDF/XML document as input and check whether the document is syntactically valid, e.g. with regards to RDF/XML. Other kinds of online validators such as URI Debugger [15] and Vapour [16] check the dereferencability of a given URI and determine whether the given URI is an information resource or a non-information resource. Other platforms, such as Jena Eyeball [17] and VRP [18], which are often used in the form of command line tools, are designed for identifying common errors in OWL or RDF documents. Generally, all of these works primarily focus on data quality problems in published datasets, and seldom provide a concrete solution for improving data quality, or attempt to identify the causes of the quality problems before the data is published. Moreover, limited attempt has been made to propose a systematic method or a set of metrics for avoiding data quality issues in LOD. In this paper, we deliberate on the importance of filtering out poor quality data by assessing the quality of a given dataset before publishing it. In the next section, our approach for proposing metrics is explained in detail.

### III. THE PROPOSED APPROACH FOR METRIC DEVELOPEMENT

The objective of our work is to identify quality deficiencies of datasets and suggest how they can be systematically evaluated before release. To this end, our approach is based on several significant quality issues identified in already published datasets on LOD. As mentioned before, only very few studies investigate the classification of such quality problems and discuss the common issues that are prevalent in the published LOD datasets [13, 14]. To the extent of our experience in publishing and interlinking academic data [7], we found that many of the published datasets suffer from quality issues such as missing values, inconsistent values and syntax errors. We believe that most of these issues have roots in the deficiencies of the sources from where that data is extracted, and they can be avoided if they are identified in the initial stages of publishing.

It is important to point out that special attention must be made to the contrast between the Closed World Assumption (CWA) versus the Open World Assumption (OWA) [6] (p. 24). The CWA is the assumption that what is not known to be true must be false, and applies when a system has complete information, e.g. database applications. The OWA is the assumption that what is not known to be true is simply unknown and applies when we want to represent knowledge (Ontologies) and discover new information [19].To exemplify, consider the following statement: *"Elena is a citizen of the USA."* Now, what if we were to ask *"Is Elena a citizen of Colombia?"* Under a CWA, the answer is *no*. Under the OWA, it is *unknown*.

As mentioned earlier, the main objective of the LOD is to crystallize knowledge through the interlinking of already existing data. According to above discussions, the LOD paradigm adheres to the OWA where generally the absence of information in a given dataset means that the information has not been made explicit and can be inferred from other available sources. Since, this work focus on the quality assessment of a dataset before interlinking to LOD, we investigate the quality deficiencies of dataset itself, not in the context of other available datasets of LOD. As a result, our approach for metric development is based on the closed world assumption. We believe that assessing quality in light of the closed world assumption holds the RDF Dataset to stricter quality standards, which is necessary when the dataset is being evaluated prior to release and in isolation. However, once validated and released as a part of the LOD, such datasets can benefit from the advantages of the open world assumption. We are only enforcing the closed world assumption in order to ensure consistency and completeness as much as possible prior to release. It is clear that this does not impact the open world assumption in any negative way.

In our work when considering which quality deficiencies to consider, the main criteria for identifying and including a quality issue was based on one of the following criteria:

The quality problems should have been spotted within published data and well documented in the literature, e.g. [13];

The existing quality issues should have been reproduced or directly observed by the authors of this paper either from first-hand experience or through observation of data on LOD;

The quality issues should be detectable and hence avoidable in the preliminary stages of data publication, i.e., prior to their publication and release to the LOD.

Therefore, our approach for metric development starts by identifying quality deficiencies of existing datasets, specifically those that can be avoided or fixed before publishing. We will then propose a set of measurement-theoretic metrics to address the identified issues, and subsequently the proposed metrics are theoretically validated and placed under empirical evaluation.

### A. PRELIMINARIES AND CONCEPTUALIZATION

Data quality assessment involves the measurement of several quality characteristics or dimensions. According to ISO 25012, a data quality characteristic (dimension) is defined by a group of data quality attributes that bear on the data quality [20]. In terms of information systems, data quality dimensions are classified into external and internal views [21]. The external view is concerned with the deployment of an information system, whereas the internal view supports a set of quality dimensions that are comparable across applications. Thus, the external view refers to why data is needed and how it is used, while in the internal view, data quality is usage-independent and can be viewed as being intrinsic to the data. Based on this classification, intrinsic quality of data is defined by being complete, unambiguous, meaningful and correct.

According to the objective of our work for identifying the quality deficiencies of datasets before release, we classify quality deficiencies into two groups: pre-publication and post-publication. By pre-publication, we mean those intrinsic quality deficiencies, which are usage-independent such as incorrectness or incompleteness. On the contrary, post-publication quality deficiencies refer to the quality problems of published datasets, which can effectively be assessed at the time of usage and not at the time of publication, e.g. timeliness that is related to the dynamicity of a dataset and depends on the time that data is actually used.

In light of the above discussion, we primarily focus on the pre-publication quality deficiencies with respect to LOD. According to the internal view of data quality, four intrinsic quality dimensions are, namely, completeness, unambiguousness, meaningfulness and correctness. In this paper, we redefine these quality dimensions in the context of LOD as *completeness, consistency, semantic accuracy* and *syntactic accuracy*.

Syntactic Accuracy: Syntactic accuracy expresses the degree to which a dataset is free of syntactic errors and refers to the valid syntax of the documents. In our work, we investigate syntactic accuracy at two levels: schema and instance levels, by detecting the erroneous representations of resources, inaccurate usage of classes and properties and misuse of RDF syntactic terms. For example, the usage of the underlying vocabularies in a given dataset is an example of quality deficiency related to syntactic accuracy at schema level.

Semantic Accuracy: Semantic accuracy relates to the correctness of a data value in comparison to the actual real world value or with the reference data agreed to be correct. In our work, resources referencing an incorrect real world correspondent and entities with erroneous attribute values are examples of quality deficiencies related to semantic accuracy.

Consistency: The consistency dimension captures the violation of semantic rules defined over (a set of) data items, where items can be tuples of relational tables or records in a file[6]. Generally, consistency implies that two or more values do not conflict with each other and it can be viewed from two perspectives, one being consistency of the same data values within a given dataset; and the other is consistency in the context of other datasets. In our work, we focus on the former and define it by the degree to which the format and the value of the data conform to the predefined schema.

Completeness: Completeness refers to the degree to which all required information is presented in a particular dataset. In this paper, there are two aspects of completeness which should be considered in CWA. The first is the schema completeness which is related to the degree to which the classes and properties are represented to describe a resource; and the second is value completeness which refers to the presence of property values based on to the schema.

Here, we will formally define key concepts and terms used throughout this paper.

- **Dataset**: An entity that consists of a schema and a set of instances, all described as RDF triples.
- **Resource**: In the Semantic Web, all real-world objects or things are called resources and are identified by URI.
- **URI:** Uniform Resource Identifier (URI) is a compact string of characters for identifying an abstract or a physical resource.
- **Instance**: A triple $t = (s, p, o)$ is an instance of a triple pattern $t_c = (s_c, p_c, o_c)$ if there exist $s_c$ is URI, $p_c$ is *rdf:type* and $o_c$ is a *class*
- **Class:** refers to a class which appears in either *o* of a triple t where p is *rdf:type*; or *s* of a triple t where p is *rdf:type* and *o* is *rdfs:Class* or *owl:Class*
- **Property:** refers to a property which appears in either *p* of a triple *t*; or *s* of a triple *t* where *p* is rdf:type and *o* is *rdf:Property*

In the following section, we investigate the quality deficiencies of RDF datasets and delineate our approach for developing appropriate metrics.

### B. IDENTIFYING QUALITY DEFICIENCIES OF A DATASET

In order to define a set of metrics for the systematic assessment of a dataset, we need to identify the common quality problems that could have been avoided prior to publication if detected. In [13], the quality problems of over 1.5 million URIs are identified in three groups: *I)* accessibility and derefencability; *ii)* syntax errors; and *iii)* noise and inconsistency. Furthermore, Furber and Hepp [14] used SPARQL queries to identify quality problems such as missing literal values or data type properties, illegal literal values, and functional dependency violations. By studying the quality issues of published data proposed in these works and based on our experience in publishing and linking academic data, we have succeeded at collecting a list of significant quality issues.

The goal was not to gather an inclusive list of issues but rather those that were mentioned primarily in the related works and also encountered during our own experience. While some of the quality deficiencies are related to the other datasets, e.g. inconsistency with other published datasets, and they are not detectable prior to the actual release of the dataset, and given the fact that the goal of our work is to assess the quality of a dataset at the initial phase of the publication process, they are not considered in this work. In the following sub-sections, we present the topology of data quality deficiencies that are only related to the dataset itself and also detectable prior to publishing.

*1. Quality deficiencies at the schema level*

In this section, we present the quality deficiencies of data at the schema level within the published datasets focusing on those which are related to the dataset itself and also detectable in the initial phase of publication. Furthermore, according to the definitions of the quality dimensions provided in Section 3.1, the most relevant quality dimensions that are affected by each of the quality issues are presented in the form of tables. In addition, the resolution method for each of the issues is provided, including validator, ontologist, domain expert, automated and/or semi-automated. Given some of the syntax errors can simply be detected by validators like Jena Eyeball[1] and VRP[2], there is no need to define metric to measure them. On the other hand, a number of quality deficiencies, such as improper usage of vocabularies, cannot be detected by validators, and also it is not possible to define systematic metrics to measure them, thus they can only be addressed with the help of human experts, i.e., either ontologists or domain experts. We are going to define metrics for those quality deficiencies that cannot be assessed either by a validator, or an expert. Table 1 enumerates the identified deficiencies at the schema level focusing on those, which are related to the dataset itself. Given the fact that the goal of our work is to assess the quality of a dataset at the initial phase of the publishing process, Table 1 presents the issues that can be detected in the phase of schema development; therefore, issues regarding extending or reusing the schema are not considered.

*Table 1. Classification of quality deficiencies at the schema level*

| Quality Deficiency | Issues | Affected quality dimensions | Resolution Method | Ref. |
|---|---|---|---|---|
| Improper usage of vocabularies | Not using appropriate existing vocabularies to describe the resources | Syntax Accuracy | Domain Experts | [1, 13] |
| Redefining existing classes/properties | Redefining the classes/properties in the ontology that already exist in the vocabularies | Consistency | Domain Experts | [1, 13] |
| Improper definition of classes/properties | - Classes with different name, but the same relations | Consistency | Semi-Automated | [7] |
| | - Properties with different name, but the same meaning | Consistency | Ontologist | [11] |
| | - Inadequate number of classes/ properties used to describe the resources | Completeness | Domain Experts | [7] |
| Misuse of data type | Not using appropriate data types for the literals | Consistency | Automated | [13, 14] |

As illustrated in Table 1, such deficiencies are classified into four groups. The first two groups are related to the concepts and terms which are already defined over the Web of Data and come from redefining instead of reusing the appropriate existing classes, properties and well-known vocabularies leading to possible redefinition of information already available on the LOD. These problems have also been investigated in [13] with the name of "ontology hijacking" when a third party redefines external classes/properties such that reasoning over data using those external terms is affected. For example, a dataset redefining the widely used core property *rdf:type* can effectively lead to every entity described on the Web being inferred as a member of this property. Given this kind of problem cannot be detected automatically; it is not possible to define systematic metrics for identifying them, thus they can only be addressed with the help of domain experts. The third group concerns issue of defining new classes and/or properties that are redundantly defined with different names or have very close resemblance and/or suffer from an inadequate number of properties/classes.

The first two issues are related to inconsistency at the schema level and consequently, affect the consistency quality dimensions of the dataset. The third issue of this group is about the amount of data provided to describe a resource, thus mostly affects the completeness aspect of a dataset. Among these issues, only redundant classes are systematically detectable, but similarity of properties and adequacy of the amount of data need more expertise to distinguish and it is very hard, if not impossible, to be automatically identified. The fourth group is related to the incorrect usage of data types, which is a relatively common error in the Web of Data affecting the consistency of published datasets. In RDF, a subset of well-defined XML data types is used to provide structure and semantics to literal values. For example, date values can be specified using the *xsd:date* data type, which provides a lexical syntax for date strings and a mapping from date strings to date values interpretable by an application[13]. Such errors can be easily identified and fixed by data publishers themselves.

---

[1] http://jena.sourceforge.net/Eyeball

[2] http://139.91.183.30:9090/RDF/VRP

### 2. Quality deficiencies at the instance level

Similar to the previous subsection, the quality deficiencies at the instance level are presented in Table 2. These instance level quality issues focus on those problems that can be detected in the initial phases of publication. Also, the most relevant quality dimensions as well as the required resolution methods for these issues are provided in the table.

Table 2. Classification of quality deficiencies at the instance level

| Quality Deficiency | Issues | Affected quality dimension | Resolution Method | Ref. |
|---|---|---|---|---|
| Errors in property values | - Missing values<br>- Out-of-range values<br>- Misspelling<br>- Inconsistent values<br>- Redundant values | Completeness<br>Semantic Accuracy<br>Semantic Accuracy<br>Consistency<br>Uniqueness | Automated<br>Automated<br>Semi-Automated<br>Automated<br>Automated | [1, 10-12, 14, 22] |
| Miss-match with the real-world | Resources without correspondence in real-world | Semantic Accuracy | Domain Experts | [2, 10, 11] |
| Syntax errors | Triples containing syntax errors | Syntactic Accuracy | Validator | [1, 13] |
| Misuse of data type/object property | Improper assignment of object property to the data type property or vice versa | Syntactic Accuracy | Validator | [13] |
| Improper usage of classes/properties | - Membership of disjoint classes<br>- Using undefined classes/properties<br>- Misplaced classes/properties | Consistency<br>Syntactic Accuracy<br>Syntactic Accuracy | Automated<br>Semi-Automated<br>Validator | [1, 13] |
| Redundant/similar individuals | Individuals with similar property values, but different names | Consistency | Ontologist | [7] |
| Invalid usage of Inverse-functional properties | Inverse-functional properties with void values | Consistency | Automated | [14] |

As shown in Table 2, seven groups of quality deficiencies are identified at the instance level that will be discussed in details in the following.

The first group is related to the errors of property values and is classified into five issues including missing values, out of range values, misspelling, inconsistent values and redundant values, each of which affect a different quality dimension as presented in this table. According to the nature of these quality issues, it is understandable that all of them are detectable automatically or semi-automatically.

The next quality deficiency in Table 2, which primarily affects semantic accuracy, is related to resources without corresponding entities within the real world. While this is a very important issue in the context of LOD, it is very hard, if not impossible, to identify with a (semi)automatic method; because, it needs domain expertise to distinguish the real-world corresponding entities of the described resources in a given dataset.

The third quality deficiency, which involves syntactic accuracy, includes the syntax errors of a dataset, e.g. mismatched data-type errors, misuse of RDF/XML shortcuts, and omission of namespace. Currently, these kinds of errors can be easily detected using validators, some of which are introduced in Section 2.

Another issue regarding syntactic accuracy is misuse of data type/object property. A data-type property describes properties, which relate some resource to a literal value, while an object property describes properties, which relate one resource to another. In some cases of published datasets, data-type properties are used between two resources or conversely, the object properties are used with literal values. This kind of errors is commonly caused by using immature APIs for producing and can easily be resolved by an appropriate syntactic validator. Similar to syntax errors, it is not necessary to define metrics for this kind of quality deficiencies.

One of the most common quality deficiencies at the instance level is the improper usage of classes and properties, including membership of disjoint classes, using undefined classes/properties and misplaced classes/properties. The first issue, which impacts dataset consistency, is related to the members of disjoint classes either asserted directly by the publisher, or inferred through reasoning. For example, the instances of classes which were defined as complements of each other (using *owl:complementOf*), or the instances of *foaf:Person* and *foaf:Document* classes in FOAF, which are defined as being disjoint. To the extent of our knowledge, there is no validator for checking this kind of inconsistency, so we will propose appropriate metrics to address it. Another quality issue refers to the usage of undefined classes/properties. In some published datasets, properties and classes are used without any formal definition. The use of ad-hoc undefined classes and properties makes automatic integration of data less effective and foregoes the possibility of making inferences through reasoning [13]. Similar to the preceding quality issue, metrics can be proposed for this that can be measured semi-automatically. Misplaced classes/properties problem is the last issue of this group, which impacts the syntactic accuracy of a dataset. This problem is related to the usage of classes as properties, or conversely the usage of properties as a class. According to the examples presented in[13], *rdfs:range* is a core RDFS property, but is defined in a document as a class. These kinds of quality issues affect syntactic accuracy of datasets, and can be avoided using an appropriate validator.

The sixth quality deficiency is related to instances that are similar in nature but are described using different terminology. Although redundancy at instance level leads to ineffective inferences through reasoning, it is not possible to identify such issues in a semi-automatic way; therefore, requires the intervention of a human expert who is familiar with ontologies.

The last deficiency presented in Table 2 refers to the invalid usage of inverse-functional properties. Aside from URIs, resources are identified by the values of properties which uniquely identify them, named "inverse-functional property"[23]. If two resources share a common value for one of these properties, reasoning will view these resources as equivalent (referring to the same resource). An example of this issue is presented in[13], where the FOAF ontology has defined *foaf:mbox* for email addresses to identify people, but there are a lot of void values for this property; and as a result all of these people are interpreted as equivalent and represent the same real-world person. The issue can easily be avoided by validating user input and also, it can automatically be resolved by checking the validity of inverse-functional values.

In summary, we have been able to identify eleven quality deficiencies characterizing nineteen quality issues at both schema and instance levels in Tables 1 and 2, respectively. Among these, six quality issues cannot be detected by any kind of automated methods and needs the intervention of human experts. It is clear that all of these metrics are very subjective and it is hard, if not impossible, to asses them automatically. The best method to measure these would be to receive experts' subjective perception using questionnaires. The remaining quality issues can be automatically detected and resolved. To the extent of our knowledge, there is no validator to cover all of these issues, particularly the issues relating to incompatibility of schema, naming, and inconsistent data. Only three quality issues can be detected directly by a validator. Thus, we propose a set of metrics to address the remaining quality issues. We note that the identified issues and the following proposed metrics are not meant to be comprehensive and are only limited to the current state of the art quality issues reported in the literature and our own practical experience with LOD datasets.

A. *Proposed Metrics*

In this section, a set of metrics are proposed to address the quality issues extracted from Tables 1 and 2 that can be resolved in an automated or semi-automated way. Bizer and Cyganiak [10] have defined a data quality assessment metric as a procedure for measuring an information quality characteristic.

Considering the fact that only a few studies have been conducted which define a set of metrics in the context of LOD[1], we need to define the required metrics with more rigors. To achieve this, metrics proposed in the areas of the Web of Data, relational databases, and data quality models have been considered[10, 20, 22, 24]; the results of which were taken into account as guidelines for designing a useful set of metrics for our purpose. Based on the mentioned quality issues, we propose ten metrics as measurement references for RDF datasets before release. The main idea behind the design of these metrics has been comprehensiveness and simplicity. It should be noted that comparable to the metrics in other domains such as software engineering, the proposed metrics may not be comprehensive, because the metrics are motivated and driven by the quality issues that were previously discussed. As a result, future research might be required to complete this set by defining new metrics from other perspectives. In this section, our proposed metrics will formally be defined. The notations used to formulate each of the metrics are presented in Table 3.

*Table 3. Notations used in formulation of the proposed metrics*

| Notation | Meaning |
|---|---|
| d | *Dataset* |
| $\mathcal{U}$ | *Set of URIs (resources) in dataset d* |
| $\mathcal{L}$ | *Set of literals* |
| Trp | *Set of triples* |
| Ins | *Set of instances in dataset d* |
| Cls | *Set of classes defined in the schema of d* |
| Prp | *Set of properties defined in the schema of d* |
| FP | *Set of functional properties in dataset d* |
| IFP | *Set of inverse functional properties in dataset d* |
| $I_{cls}$ | *Set of instances of class 'cls' in dataset d* |
| $T_{prp}$ | *Set of triples using property 'prp' as predicate* |
| $T_{cls}$ | *Set of triples using class 'cls' as object* |
| Dom (x) | Domain of x as defined in the schema of d |
| Rng (x) | Range of x as defined in the schema of d |
| $x \not\equiv y$ | x is disjoint with y |
| t.o | Object of triple t |
| t.s | Subject of triple t |
| t.p | Predicate of triple t |
| x.name | Name of x |
| x.type | Type of x (for object property, it is URI, and for data type property, is data type as defined in the schema) |
| x.vlu | *Value of x* |
| Valid | Set of terms in a given dictionary |

For more clarity, we have developed a hypothetical ontology called Family using Protégé to exemplify each of the metrics definitions as shown in Fig. 1, there are 18 classes in the Family ontology such as 'Sex', 'Person', 'Father' …, and 7 instances are defined for these classes, e.g. 'Math' and 'Peter'.

The number of properties in the Family ontology is 17 including 11 object properties, such as 'hasSex' and 'isMotherOf', and 6 data type properties, e.g. 'hasFirstName' and 'hasBirthYear'. In the following subsections, each of the metrics is formally defined and exemplified using the Family ontology.

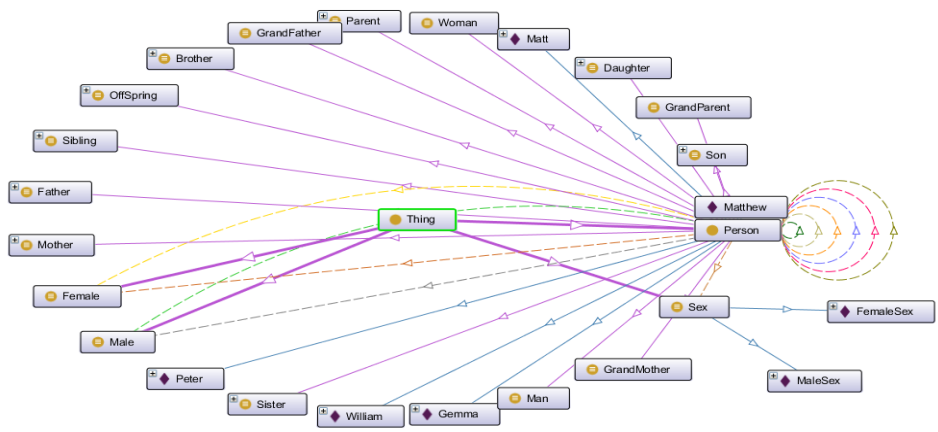

*Fig. 1. Hypothetical ontology*

*Missing Property Values (Miss_Vlu)*

The Missing Property Values metric measures the ratio of the properties defined in the schema of dataset d, but not presented in d. Miss_Vlu is calculated as:

$$Miss\_Vlu_d = 1 - \frac{\sum_{\forall p \in Prp} |T_p|}{|Cls|*|Prp|} \quad (1)$$

Through Miss_Vlu, we measure the presence of required properties for each instance according to the defined properties in the schema. Based on Equation (1), Tp is the number of uses of a specific property p in the dataset, thus, sum of the value of Tp should be computed. In our example, there are 37 triples that use the defined properties in the Family ontology. Also, regarding the number of properties and classes, the value of |Cls|*|Prp| is equal to 306 (18*17). As a result, the value of Miss_Prp metric is 88% (1- 37/306).

Out-of-range Property Values (Out_Vlu)

The Out_of_range Property Values metric measures the ratio of the triples of dataset d which contain properties with out of range values. Out_Vlu is calculated as:

$$Out\_Vlu_d = \frac{\sum_{\forall t \in Trp} Out(t)}{|Trp|} \quad (2)$$

Where: $Out(t) =$
$\begin{cases} 1 & if\ (t.o \in Cls) \land (t.o.vlu \neq Rng(t.p)) \\ & or\ (t.o \in \mathcal{L}) \land (t.o.vlu \neq Rng(t.o)) \\ 0 & otherwise \end{cases}$

Based on Equation (2), Out_Vlu measures the ratio of triples containing out of range properties, both data type properties and object properties. Given that <family:hasSibiling> is an object property with the range of <family:Person> and Ali and Sara are instances of <family.Person>. Also, <family:MaleSex> is an instance of <family:Sex>. We can create this type of quality error by changing a triple as follows:

<family:Math> <family:hasSibling> <family:Gemma> into

<family:Math> <family:hasSibling> <family:MaleSex>

By this change, the number of triples containing out of range properties is increased by one and the numerator of fraction of Out_Vlu metric is increased by one as well.

Misspelled Property Values (Msspl_Prp_Vlu)

The Misspelled Property Values metric measures the ratio of the properties of dataset d which contain misspelled values. Msspl_Prp_Vlu is calculated as:

$$Msspl\_Prp\_Vlu_d = \frac{\sum_{\forall t \in Trp} Msp(t)}{|Trp|} \quad (3)$$

Where: $Msp(t) =$
$\begin{cases} 1 & if\ (t.o \in \mathcal{L}) \land (t.o.vlu \notin Valid) \\ 0 & otherwise \end{cases}$

Msspl_Prp_Vlu is defined to measure the misspelling errors of the values of data type properties. To this end, we have used Lucene spell checker [25] in our implementation. This spell checker includes different languages, e.g. English, Danish, Dutch and Spanish. Given that <family:hasFamilyName> is a data type property and 'Smith' is a valid term in our dictionary, but 'Smithp' is not. We can insert a misspelling error in a triple as follows:

< family:Math> <family:hasFamilyName> < family:Smithp>

Similar to Out_Vlu metric, by this change, numerator of fraction of Msspl_Prp_Vlu metric is increased by one.

Undefined Classes and Properties (Und_Cls_Prp)

The undefined classes and properties metric measures the ratio of the triples of dataset d using classes or properties without any formal definition. Und_Cls_Prp is calculated as:

$$Und\_Cls\_Prp_d = \frac{\sum_{\forall t \in Trp}[Undc(t.c)+Undp(t.p)]}{|Trp|} \quad (4)$$

Where:

$Undc(t.c) = \begin{cases} 1 & if\ (t \in T_c) \land (c \notin Cls) \\ 0 & otherwise \end{cases}$

$Undp(t.p) = \begin{cases} 1 & if\ (t \in T_p) \land (p \notin Prp) \\ 0 & otherwise \end{cases}$

Und_Cls_Prp is defined to detect the classes and properties used, but not defined in the schema. For example, consider the following triple:

< family:Ali> <rdf:type> <family:human>

Where human is a class that does not exist in the schema. In this case, Und_Cls_Prp metric is increased.

Membership of Disjoint Classes (Dsj_Cls)

The Membership of Disjoint Classes metric measures the ratio of the instances of dataset d being members of disjoint classes. Dsj_Cls is calculated as:

$$Dsj\_Cls_d = \frac{\sum_{\forall i \in Ins} Dsjc(i)}{|Ins|} \quad (5)$$

Where: $Dsjc(i) =$
$\begin{cases} 1 & if\ (i \in \{I_{c1} \cap I_{c2}\}) \land (c1 \not\equiv c2) \\ 0 & otherwise \end{cases}$

Based on Equation (5), it is understood that the value of Dsj_Cls metric is increased by defining an individual as instance of two disjoint classes. For example, consider the following triples:

<family:Ali> <rdf:type> <family:Female>

<family:Ali> <rdf:type> <family:Male>

When both of these triples exist in a dataset, the numerator of Dsj_Cls metric will be increased by one.

Inconsistent Property Values (Inc_Prp_Vlu)

The Inconsistent Property Values metric measures the ratio of the triples of dataset d in which the values of properties are inconsistent. Inc_Prp_Vlu is calculated as:

$$\text{Inc\_Prp\_Vlu}_d = \frac{\sum_{\forall t \in Trp} IPV(t)}{|Trp|} \qquad (6)$$

Where:
$$IPV(t) = \begin{cases} 1 & \forall t(s.p.o) \in Trp. \exists (t'(s.p.o')\in Trp) \mid (o.type \neq o'.type) \wedge (o.vlu \neq o'.vlu) \\ 0 & otherwise \end{cases}$$

According to Equation (6), Inc_Prp_Vlu counts the triples in which the subjects and predicates are the same, but their objects are different in term of both values and types, e.g.

<family:Ali> <family:hasSibiling> <"Sara">

<family:Ali> <family:hasSibiling> <family:Sara>

When both of these triples exist in a dataset, the numerator of Inc_Prp_Vlu metric will be increased by one.

Functional Properties with Inconsistent Values (FP)

The FP metric measures the ratio of the number of triples of dataset d with functional properties which contain inconsistent values. It is calculated as:

$$FP_d = \frac{\sum_{\forall t \in Trp} FP(t)}{|Trp|} \qquad (7)$$

Where:
$$FP(t) = \begin{cases} 1 & \forall t(s.p.o) \in Trp \wedge (p \in FP). \exists t'(s.p.o') \in Trp \mid (o \neq o') \\ 0 & otherwise \end{cases}$$

According to this definition, FP computes the number of triples whose predicates are a specific functional property with the same subjects, but different objects. Given <family:hasMother> is a functional property and both of following triples exist in our dataset:

<family:Ali> <family:hasMother> <family:Mari>

<family:Ali> <family:hasMother> <family:Sara>

In this case, the numerator of FP metric will be increased by one.

Invalid Usage of Inverse Functional Properties (IFP)

This metric measures the ratio of the number of triples of dataset d which contain invalid usage of inverse-functional properties. IFP is calculated as:

$$IFP_d = \frac{\sum_{\forall t \in Trp} IFP(t)}{|Trp|} \qquad (8)$$

Where:
$$IFP(t) = \begin{cases} 1 & \forall t(s.p.o) \in Trp \wedge (p \in IFP). \exists t'(s'.p.o) \in Trp \mid (s \neq s') \\ 0 & otherwise \end{cases}$$

The definition of IFP is similar to FP, where IFP counts the triples in which their predicates have the same inverse functional property with the same objects, but different subjects. For example, given <family:isMotherOf> is an inverse functional property and both of the following triples exist in our dataset:

<family:Mari> <family:isMotherOf> <family:Ali>

<family:Sara> <family:isMotherOf> <family:Ali>

In this case, we expect the numerator of IFP metric to increase by one.

*Improper Data Type for Literal (Im_DT)*

The Improper Data Type for Literal metric measures the ratio of the number of triples of dataset d which contain data type properties with inappropriate data types. Im_DT is calculated as:

$$\boldsymbol{Im\_DT_d} = \frac{\sum_{\forall t \in Trp} IDT(t)}{|Trp|} \qquad (9)$$

Where:
$$IDT(t) = \begin{cases} 1 & if[(t.o \in L) \wedge (t.o.type \neq Rng(t.p))] \\ 0 & otherwise \end{cases}$$

For example, if the range of a data type property, e.g. <family:hasBirthYear>, is defined as an integer and the following triple exists in our dataset:

<family:Gemma> < family:hasBirthYear> <'1996'>

Then, the numerator of Im_DT metric is increased by one.

Similar Classes (Sml_Cls)

The Similar Classes metric measures the ratio of the classes of dataset d with different names, but the same instances. Sml_Cls is calculated as:

$$\boldsymbol{Sml\_Cls_d} = \frac{\sum_{\forall c \in Cls} SC(c)}{|Cls|} \qquad (10)$$

Where:
$$SC(c) = \begin{cases} 1 & \forall c \in Cls, \exists (c' \in Cls) \mid (c' \nsubseteq c) \wedge (c \nsubseteq c') \wedge (I_c \equiv I_{c'}) \wedge (c.name \neq c'.name) \\ 0 & otherwise \end{cases}$$

For example, given we have included another class called <family:human>, in which the set of instances of <family:person> is equivalent to the set of instances of <family:human>. Then, <family:person> and <family:human> classes are considered as similar. According to the definitions presented for the metrics, it is clear that all of the metrics are defined to measure the quality problems in the scope of the RDF dataset itself, not in the context of other datasets.

We have tried to cover as many deficiencies as possible that can be identified prior to the publication of a dataset, which is the focus of our work.

From the level of quality deficiency point of view, we have addressed the quality issues of a dataset at both schema level and instance level.

As mentioned earlier, the last two metrics (M9 and M10) are defined to address intrinsic quality issues at the schema level, while the others are related to the intrinsic quality problems at the instance level. According to [26], the preferred way for metric definition is to calculate the number of the undesirable outcomes divided by that of the total outcomes.

Thus, all of the formulas presented for computation of quality deficiencies illustrate the undesirable outcomes using the ratio scale.

## IV. EMPIRICAL EVALUATION

The main purpose of our work is to propose a set of appropriate metrics to address the quality issues of RDF datasets before their publication. For this purpose, it is necessary to place them under empirical evaluation to observe their behavior and show their applicability in practice.

Hence, we first calculated the values of the metrics for eight datasets in order to show the metric behavior over datasets of different domains and sizes. Next, we manipulated the quality of these datasets by applying some heuristics, and then recalculated the metric values to observe the behavior of the metrics over these changes. In the following subsections, the results of these observations are presented.

## V. FIRST OBSERVATION OVER THE ORIGINAL DATASETS

In this section, we report the results of our first observations with regards to the calculation of the proposed metrics for several real world datasets.

We have selected eight datasets from the EU FP6 Networked Ontology (NeOn) project [27] with the restriction that the language of the datasets needed to be English.

Also, we made sure that these datasets: I) were from across variety of domains; and ii) were of different sizes in terms of the number of triples in the datasets as shown in Table 4.

*Table 4. The details of the datasets used in the first experiment*

| No. | Original Datasets | No. of triples | No. of instances | No. of classes | No. of properties |
|---|---|---|---|---|---|
| 1 | FAO Water Areas | 5,365 | 293 | 7 | 19 |
| 2 | Water Economic Zones | 25,959 | 693 | 22 | 127 |
| 3 | Large Marine Ecosystems | 6,006 | 358 | 9 | 31 |
| 4 | Geopolitical Entities | 22,725 | 312 | 11 | 101 |
| 5 | ISSCAAP Species Classification | 368,619 | 23,856 | 22 | 93 |
| 6 | Species Taxonomic Classification | 318,153 | 11,738 | 5 | 26 |
| 7 | Commodities | 28,210 | 1,394 | 6 | 19 |
| 8 | Vessels | 2,118 | 120 | 6 | 22 |

The calculation of the values for each metric was done automatically. We have implemented an automated tool that is able to automatically compute the metric values for any given input dataset. The code of the implemented tool as well as the employed datasets are available publicly [28]. Table 5 presents all of the collected values for the ten proposed metrics for each of the original datasets. In light of the values of the metrics as reported in Table5, it is clear that four of the proposed metrics have the same value regardless of the dataset. These metrics are M4(Und_Cls_Prp), M5(Dsj_Cls), M7(FP) and M8(IFP). In all cases, the value of '0' indicates that such deficiencies do not exist in any of the subject datasets. The most likely reason is that an automated tool can easily identify and resolve such redundancies and therefore data publishers have most likely already resolved them in the eight published datasets that we used.

In light of the fact that the goal of proposed metric is to assess a dataset before release, and we would like to ensure the suitability of all proposed metrics in practice, in the following section, we will systematically reduce the quality of the datasets and recalculate the values of the metrics over the modified datasets.

Table 5. Results of first experiment

| Datasets | M1 | M2 | M3 | M4 | M5 | M6 | M7 | M8 | M9 | M10 |
|---|---|---|---|---|---|---|---|---|---|---|
| FAO Water Areas | 0.33 | 0.16 | 0.16 | 0.00 | 0.00 | 0.20 | 0.00 | 0.00 | 0.00 | 0.22 |
| Water Economic Zones | 0.74 | 0.19 | 0.00 | 0.00 | 0.00 | 0.19 | 0.00 | 0.00 | 0.00 | 0.77 |
| Large Marine Ecosystems | 0.56 | 0.22 | 0.15 | 0.00 | 0.00 | 0.19 | 0.00 | 0.00 | 0.00 | 0.11 |
| Geopolitical Entities | 0.41 | 0.22 | 0.00 | 0.00 | 0.00 | 0.19 | 0.00 | 0.00 | 0.00 | 0.55 |
| ISSCAAP Species Classification | 0.85 | 0.88 | 0.05 | 0.00 | 0.00 | 0.32 | 0.00 | 0.00 | 0.43 | 0.05 |
| Species Taxonomic Classification | 0.38 | 0.96 | 0.05 | 0.00 | 0.00 | 0.33 | 0.00 | 0.00 | 0.48 | 0.00 |
| Commodities | 0.30 | 0.30 | 0.15 | 0.00 | 0.00 | 0.37 | 0.00 | 0.00 | 0.00 | 1.00 |
| Vessels | 0.33 | 0.89 | 0.17 | 0.00 | 0.00 | 0.17 | 0.00 | 0.00 | 0.00 | 0.50 |
| Mean | 0.49 | 0.48 | 0.09 | 0.00 | 0.00 | 0.25 | 0.00 | 0.00 | 0.11 | 0.40 |
| STDEV | -0.21 | -0.36 | -0.07 | 0.00 | 0.00 | -0.08 | 0.00 | 0.00 | -0.21 | -0.36 |

## VI. MANIPULATING THE DATASETS

In order to understand the behavior of the proposed metrics over different datasets with diverse set of quality issues, we systematically reduced the quality of the datasets and recalculated the values of the metrics over the modified datasets. As a result, we were able to show the changing trend of the proposed metrics over datasets of the same nature but with different quality issues, i.e., how the values for the metrics changed if quality deficiencies were introduced into the same datasets. For this purpose, we contaminated the datasets using some heuristics to ensure that all of the mentioned quality issues are present in the eight datasets. The applied heuristics for creating and injecting each of the quality issues are introduced in Table 6.

Table 6. Heuristics for manipulation

| Quality Issues | Heuristic | Level |
|---|---|---|
| Missing Property Values | H1 - Creating new properties at the schema<br>H2 - Removing triples randomly | Schema<br>Instance |
| Out-of-range Property Values | H3 - Assigning out of range values to the data type properties | Instance |
| Misspelled Property Values | H4 - Removing or inserting some characters to the literals<br>H5 - Replacing a literal by one not included in the dictionary | Instance |
| Undefined Class and Properties | H6- Renaming the properties and classes in a number of triples<br>H7- Removing the definitions of classes and properties | Instance<br>Schema |
| Membership of Disjoint Classes | H8- Making classes with common instances as disjoint classes<br>H9- Creating new instances as member of two disjoint classes | Schema<br>Instance |
| Inconsistent Property Values | H10- Creating new triples with inconsistent property values | Instance |
| Functional Properties with Inconsistent Values | H11- Identifying the triples containing functional properties and copying those triples with different objects | Instance |
| Invalid Usage of Inverse Functional Properties | H12- Identifying the triples containing inverse functional properties and copying those triples with different subjects | Instance |
| Improper Data Type for Literal | H13- Changing data type of a number of data type properties | Schema |
| Similar Classes | H14- Copying some classes with different names and creating the same instances for copied classes | Schema |

As shown in Table 6, fourteen heuristics are applied within the dataset contamination process. Some of these quality issues such as misspelling errors were made using an ontology editor, i.e. Protégé. For those quality issues, such as invalid usage of inverse functional properties, errors were introduced manually. The distribution model for applying heuristics over the datasets is presented in Table 7.

*Table 7. Distribution of heuristics over datasets*

| Datasets | H1 | H2 | H3 | H4 | H5 | H6 | H7 | H8 | H9 | H10 | H11 | H12 | H13 | H14 |
|---|---|---|---|---|---|---|---|---|---|---|---|---|---|---|
| FAO Water Areas | 9 | | 150 | 12 | | | 5 | | | 450 | | 10 | 205 | 7 |
| Water Economic Zones | 3 | | | 50 | | | | 4 | 6 | | 12 | | | 1 |
| Large Marine Ecosystems | 9 | | | | | 150 | 2 | | | | | 10 | | 2 |
| Geopolitical Entities | 2 | | | | 75 | | 4 | | | 209 | | | | 7 |
| ISSCAAP Species Classification | 8 | 10 | | 150 | | | | | | | | | | 1 |
| Species Taxonomic Classification | 1 | 3500 | 40 | | | | | | | | | | | 1 |
| Commodities | 7 | 15 | | | | | | | | | | 20 | | 2 |
| Vessels | 3 | 20 | | | 10 | | | | | | | 20 | | 1 |

As seen in Table 7, we have randomly applied the heuristics to the different datasets. The rationale for this was to measure the values for our metrics both before and after the quality issues were injected. In the next section, the values of all metrics after the application of the heuristics are presented.

VII. SECOND OBSERVATION OVER THE MANIPULATED DATASETS

Following our approach for investigating the behavior of the metrics over the datasets with different quality issues, the results of the second experiments are presented in this section. The aim of the second experiment is to show the trends of proposed metrics by recalculating the values of the metrics over manipulated datasets. As a result, we expect to observe meaningful changes in the values of the metrics according to the heuristics used to manipulate the datasets. The manipulated datasets are publicly available [29]. Table 8 presents the details of experimented datasets after manipulation.

Based on the application of the contamination heuristics on the datasets, the size of our datasets could have changed in terms of numbers of triples, instances, classes and properties. This is the direct result of the application of the heuristics on the datasets. For instance, there are two ways for creating missing properties: I) creating a new property that does not have any corresponding instances using it, and ii) removing some of the existing property values of existing instances. The former causes an increase in terms of the numbers of triples, instances and properties, while the latter reduces the number of triples. As a result of the applied heuristics, the size of the subject datasets is changed as shown in Table 8.

*Table 8. The details of the datasets used in the second experiment*

| No. | Contaminated Datasets | No. of triples | No. of instances | No. of classes | No. of properties |
|---|---|---|---|---|---|
| 1 | Dirty FAO Water Areas | 6,025 | 293 | 9 | 28 |
| 2 | Dirty Water Economic Zones | 25,999 | 693 | 23 | 130 |
| 3 | Dirty Large Marine Ecosystems | 6,018 | 358 | 11 | 40 |
| 4 | Dirty Geopolitical Entities | 23,037 | 312 | 14 | 103 |
| 5 | Dirty ISSCAAP Species Classification | 368,619 | 23,856 | 23 | 101 |
| 6 | Dirty Species Taxonomic Classification | 314,628 | 11,738 | 6 | 27 |
| 7 | Dirty Commodities | 28,210 | 1,394 | 8 | 26 |
| 8 | Dirty Vessels | 2,092 | 122 | 7 | 25 |

In the same way as in the first experiment, the values for all of the proposed metrics were calculated using our implemented tool that is able to automatically compute the metric values for any given input dataset. Table 9 presents all of the collected values of ten proposed metrics for each of the contaminated datasets as well as the amount it was changed compared to the values in Table 5 (shown by Δ).

*Table 9. Results of second experiment*

| Contaminated Datasets | M1 | M2 | M3 | M4 | M5 | M6 | M7 | M8 | M9 | M10 |
|---|---|---|---|---|---|---|---|---|---|---|
| Dirty FAO Water Areas | 0.56 | 0.20 | 0.16 | 0.00 | 0.00 | 0.28 | 0.00 | 0.00 | 0.05 | 0.29 |
| Δ | 0.23 | 0.03 | - | - | - | 0.09 | - | - | 0.05 | 0.07 |
| Dirty Water Economic Zones | 0.74 | 0.19 | 0.00 | 0.00 | 0.01 | 0.19 | 0.00 | 0.00 | 0.00 | 0.78 |
| Δ | 0.01 | - | - | - | 0.01 | - | - | - | - | 0.01 |
| Dirty Large Marine Ecosystems | 0.69 | 0.22 | 0.15 | 0.03 | 0.00 | 0.19 | 0.00 | 0.00 | 0.00 | 0.09 |
| Δ | 0.13 | - | - | 0.03 | - | - | - | - | - | -0.02 |
| Dirty Geopolitical Entities | 0.42 | 0.22 | 0.00 | 0.00 | 0.00 | 0.20 | 0.00 | 0.00 | 0.00 | 0.07 |
| Δ | 0.01 | - | - | - | - | 0.01 | - | - | - | -0.47 |
| Dirty ISSCAAP Species Classification | 0.87 | 0.88 | 0.05 | 0.00 | 0.00 | 0.32 | 0.00 | 0.00 | 0.43 | 0.05 |
| Δ | 0.02 | - | - | - | - | - | - | - | - | - |
| Dirty Species Taxonomic Classification | 0.44 | 0.96 | 0.05 | 0.00 | 0.00 | 0.36 | 0.03 | 0.03 | 0.47 | 0.00 |
| Δ | 0.06 | - | - | - | - | 0.03 | 0.03 | - | -0.01 | - |
| Dirty Commodities | 0.58 | 0.30 | 0.15 | 0.00 | 0.00 | 0.37 | 0.05 | 0.00 | 0.00 | 0.88 |
| Δ | 0.28 | - | - | - | - | - | 0.05 | - | - | -0.13 |
| Dirty Vessels | 0.47 | 0.88 | 0.14 | 0.00 | 0.00 | 0.17 | 0.00 | 0.01 | 0.00 | 0.57 |
| Δ | 0.13 | - | -0.02 | - | - | - | - | 0.01 | - | 0.07 |

Table 9 illustrates the result of applying metrics over the contaminated datasets. Given the fact that the goal of the proposed metrics is to assess a dataset before its release and in light of the fact that the datasets of this observation are manually contaminated, it is conceivable that some of the quality issues will impact the others. In the next section, these values will be discussed in detail.

VIII. STATISTICAL ANALYSIS AND DISCUSSION

In this section, the suitability of the proposed metrics will be investigated. To this end, first the trends of metrics over two sets of observations are discussed. Then, in order to study metric interdependencies, the Spearman's Rho correlation is used to show whether some of the metrics capture similar aspects of the datasets and whether they are overlapping or not.

*B. The Trends of Metrics over two Sets of Observations*

In this subsection, we investigate the trends of the metric values over two sets of observations and show the relationship between heuristics and metric values. We expect that all of the metrics are properly changed after contamination. For better comparison of the results, we present both the amount of data contamination and the values of the metrics in the same table, as seen in Table 10.

Table 10 shows that there are two rows dedicated to each dataset. The first row shows the numbers of heuristics applied to a dataset and the second row is the change of metric values after the application of the heuristic(s). Also, the columns of this table depict the heuristics which are grouped based on ten corresponding quality issues presented in Table 6. Because of the difference in the nature of contaminations, the results of the heuristics in the same group of quality problems are not additive; as a result, they are presented separately, e.g. H1+H2. This way, we can simply compare the effect of heuristics with the corresponding metric values. By applying heuristics over datasets, the metric values have reacted in three ways: increasing, decreasing, not changing, as the result of Δ metrics can be positive, negative or zero. In addition, after applying heuristics, we have observed a few cases that metric values are changed without applying corresponding metrics, e.g. ΔM6 and ΔM7 for Species Taxonomic Classification. For better discussion on different states of metric values and heuristics, the possible states are shown in Table 11.

*Table 10. Trends of Metrics over Manipulation*

| Heuristics | H1+H2 | H3 | H4+H5 | H6+H7 | H8+H9 | H10 | H11 | H12 | H13 | H14 |
|---|---|---|---|---|---|---|---|---|---|---|
| Δ Metrics | Δ M1 | Δ M2 | Δ M3 | Δ M4 | Δ M5 | Δ M6 | Δ M7 | Δ M8 | Δ M9 | Δ M10 |
| Dirty FAO Water Areas | 9+0 | 150 | 12+0 | 0+5 | 0 | 450 | 0 | 10 | 205 | 7 |
|  | 0.23 | 0.03 | 0 | 0 | 0 | 0.09 | 0 | 0 | 0.05 | 0.07 |
| Water Economic Zones | 3+0 | 0 | 50+0 | 0 | 4+6 | 0 | 12 | 0 | 0 | 1 |
|  | 0.01 | 0 | 0 | 0 | 0.01 | 0 | 0 | 0 | 0 | 0.01 |
| Large Marine Ecosystems | 9+0 | 0 | 0 | 150+2 | 0 | 0 | 0 | 10 | 0 | 2 |
|  | 0.13 | 0 | 0 | 0.03 | 0 | 0 | 0 | 0 | 0 | -0.02 |
| Geopolitical Entities | 2+0 | 0 | 0+75 | 0+4 | 0 | 209 | 0 | 0 | 0 | 7 |
|  | 0.01 | 0 | 0 | 0 | 0 | 0.01 | 0 | 0 | 0 | -0.47 |
| ISSCAAP Species Classification | 8+10 | 0 | 150+0 | 0 | 0 | 0 | 0 | 0 | 0 | 1 |
|  | 0.02 | 0 | 0 | 0 | 0 | 0 | 0 | 0 | 0 | 0 |
| Species Taxonomic Classification | 1+3500 | 40 | 0 | 0 | 0 | 0 | 0 | 0 | 0 | 1 |
|  | 0.06 | 0 | 0 | 0 | 0 | 0.03 | 0.03 | 0 | -0.01 | 0 |
| Commodities | 7+15 | 0 | 0 | 0 | 0 | 0 | 20 | 0 | 0 | 2 |
|  | 0.28 | 0 | 0 | 0 | 0 | 0 | 0.05 | 0 | 0 | -0.13 |
| Vessels | 3+20 | 0 | 0+10 | 0 | 0 | 0 | 0 | 20 | 0 | 1 |
|  | 0.13 | 0 | -0.02 | 0 | 0 | 0 | 0 | 0.01 | 0 | 0.07 |

*Table 11: The rate of Metric changes with the heuristics*

|  |  | Heuristics | |
|---|---|---|---|
|  |  | Applied | Not applied |
| Δ Metrics | Positive | 23% | 3% |
|  | Negative | 5% | 1% |
|  | Zero | 14% | 55% |

As mentioned earlier, in this experiment, we would ideally expect the metric values to increase only by applying corresponding heuristic(s). In light of the values reported in Table 11, it is observed that most of the outcomes are desirable (23%+55%); however, some of the values need more discussions which are presented as follows.

- *Heuristics have been applied and changes of corresponding metrics are positive*

As shown in the first column of Table 10, it can be seen that M1 (Miss_Vlu) has a normal reaction to the corresponding heuristics (H1+H2), since it has increased with respect to the different sizes of the datasets. For example, consider (H1+H2) on Species Taxonomic Classification dataset with 314,628 triples. The values of 'H1+H2' are '1+3500' and M1 has positively changed by 0.06; while when '7+15' (H1+H2) are applied on Commodities with 28,210 triples, M1 has increased by 0.28. By comparing two values of '0.06' and '0.28', we observe that ΔM1 in the latter is more that ΔM1 in the former, while the number of H2 in the former is much more than the latter. This has occurred because of the different sizes of these datasets. Thus, we can conclude that ΔM1 is related to the ratio of the number of affected triples over the total number of triples in the dataset. According to the information reported in Table 11, 23% of the values reported in Table 10, are in this category and it means that in these cases, the metrics react as expected to the corresponding heuristics.

- *Heuristics have been applied and changes of corresponding metrics are negative*

As reported in the last column of Table 10, we have observed that by applying heuristic (H14), the rate of change of M10 (Sml_Cls) is negative. In other word, by copying some classes with different names, metric M10 (Sml_Cls), which is defined to measure the similar classes, reacts worse. This happens whenever some other heuristics are manipulating the classes, e.g. H6 and H7. Thus, when these heuristics are concurrently applied on a dataset, the effects of applying H6+H7 are more than the effect of H14. This also applies to the effects of H3 on H13 as both of these heuristics are manipulating data type properties.

- *Heuristics have been applied and corresponding metrics have not changed*

In light of the values reported in Table 11, in 14% of the results the metric values have not changed, when the corresponding heuristics were applied. There are two reasons for this. The first is related to the side effects of applying dependent heuristics on the same dataset, which is described earlier, and the second reason comes from the precision of reported values, i.e. 0.01. For example, by applying H4 on the ISSCAAP Species Classification dataset, we have created misspelled property values by changing the literals used in 150 triples, and we expect the corresponding metric (M3), which measures the ratio of triples containing misspelled property values, to change, but we observe that ΔM3 is zero. The reason is that the change of M3 is less than 1% of triples, and as a result it cannot be shown in the table, because all of the values are presented with the precision of 0.01. This reason is applied to all similar situations that the rate of change of data is less than 1%.

- *The metric values have changed (positive/negative) without applying corresponding heuristics*

According to Table 10, three cases are reported for Species Taxonomic Classification dataset, in which metric values have changed without applying corresponding heuristics, two of which indicate positive change and one is negative. The positive changes occurred when H10 and H11 have been applied, which are respectively corresponding to M6 (Inc_Prp_Vlu) and M7 (FP). M6 computes the ratio of occurrence of inconsistent property values, while M7 measures the invalid usage of functional properties in triples. Based on the formulas presented in Section 3.3, both of these metrics are calculating the ratio of quality issues by dividing the undesirable triples to the total number of triples. Since the number of triples has changed after applying H2 on the mentioned datasets, the values of metrics M6 and M7 also changed, because the denominator of the fraction in both are the number of triples. Also, the negative value of '-0.01' for M9 is related to the effect of applying H3 over Species Taxonomic Classification dataset, for the same reason mentioned in 5.1.2.

- *Heuristics have not been applied and the metric values have not changed*

Regarding Table 10, it can be realized that more than half of the reported values refer to these cases in which neither heuristic, nor changes of metrics are observed. In light of above discussion and regarding to the information provided in Table 11, we can conclude that in 78% (23%+55%) of the scenarios the metrics have behaved as expected.

### C.  Threats to the validity of observations

The aim of our dataset manipulation work was to investigate the trends of metrics over real datasets and to compare the results of applying metrics on good and poor quality data. After manipulating the datasets, we have observed some remarkable points which are summarized as follows:

- Some of the heuristics were not independent and as a result, the order of applying these heuristics affected the results of measuring the quality problems by the metrics. This occurs because introducing some errors into a given dataset can have a number of side effects on other metrics. For example, when a given property of 'gender' is removed from the ontology in order to create undefined properties in the dataset, we would only expect increase in M4 (Und_Cls_Prp), but the value of M1 (Miss_Vlu) would change as well as a side effect. The reason is that all of the instances that have not used the 'gender' property, are already calculated by the first metric as "missing property values", and by removing the 'gender' property, those instances will not be taken into consideration any longer when computing the first metric. Thus, the change in one quality issue can implicitly impact other quality issues and therefore, their corresponding metrics. For better investigation of metrics behavior, it is better to not concurrently apply these heuristics on the same dataset.
- Another factor affecting our results was related to precision of reported values, i.e. 0.01. Based on this experiment, whenever the number of changes is less than 1% of the number of triples, the changes of metric values cannot properly be reported, e.g. applying H4 on ISSCAAP Species Classification in our observation as explained in Section 5.1.3.
- Although in our scenario, no radical shift in the metric values was observed, but we are not going to generalize our finding about the trends of metrics, because of the limited number of datasets that we have used in this experiment. As a result, we believe more experiments need to be done to reach a valid conclusion about the reaction (both positive and negative) of metrics to the changes in the measurement subject.

To further investigate metrics' behaviors, we empirically study the interdependency of metrics using a Spearman's Rho correlation test in the next section.

D. Metrics Interdependency Study

In this subsection, we investigate the interdependency of metrics and perform Spearman's Rho correlation test to measure inter-metric correlation. In other words, the goal of this study is to see whether some of the metrics capture similar aspects of the datasets and whether they are overlapping or not. Ideally, we would like each metric to represent a distinct aspect of the quality issues in the datasets. The results of this study are shown in Table 12.

*Table 12: Inter-metric correlation*

| Rho | M1 | M2 | M3 | M4 | M5 | M6 | M7 | M8 | M9 | M10 |
|---|---|---|---|---|---|---|---|---|---|---|
| M1 |  | -0.24 | 0.08 | 0.25 | 0.41 | -0.01 | -0.20 | -0.41 | 0.00 | 0.24 |
| p value |  | - | - | - | - | - | - | - | - | - |
| M2 |  |  | -0.02 | -0.17 | -0.58 | 0.36 | 0.46 | 0.58 | 0.50 | -0.49 |
| p value |  |  | - | - | - | - | - | - | - | - |
| M3 |  |  |  | 0.34 | -0.50 | 0.16 | 0.17 | -0.17 | 0.04 | 0.28 |
| p value |  |  |  | - | - | - | - | - | - | - |
| M4 |  |  |  |  | -0.14 | -0.33 | -0.22 | -0.14 | -0.28 | -0.08 |
| p value |  |  |  |  | - | - | - | - | - | - |
| M5 |  |  |  |  |  | -0.33 | -0.22 | -0.14 | -0.28 | 0.41 |
| p value |  |  |  |  |  | - | - | - | - | - |
| M6 |  |  |  |  |  |  | 0.77 | 0.41 | 0.55 | -0.20 |
| p value |  |  |  |  |  |  | * | - | - | - |
| M7 |  |  |  |  |  |  |  | 0.54 | 0.20 | 0.11 |
| p value |  |  |  |  |  |  |  | - | - | - |
| M8 |  |  |  |  |  |  |  |  | 0.66 | -0.58 |
| p value |  |  |  |  |  |  |  |  | * | - |
| M9 |  |  |  |  |  |  |  |  |  | -0.71 |
| p value |  |  |  |  |  |  |  |  |  | * |
| M10 |  |  |  |  |  |  |  |  |  |  |
| p value |  |  |  |  |  |  |  |  |  |  |

('-' means p value>0.05 and '*' means p value ≤ 0.05)

In Table 12, there are two rows dedicated to each metric; the values on the upper rows show the degree of correlation (rho value), while the symbols on the lower rows depict the significance of the correlation (p value). As mentioned earlier, a significance level of $p < 0.05$ is used to accept the results of the correlation. According to Spearman's correlation, a correlation with a significance p value ≤ 0.05 can be considered to be significant. Therefore, in our work such correlations are depicted with the symbol of '*' and exhibit meaningful correlation between metrics; while the other symbol of '-' between each pair of the metrics shows that those metrics are independent.

The correlations values found between the metrics show that only three values are depicted with "*", which means that their corresponding p-values are less than 0.05, while most of the proposed metrics are in fact not dependent. There are three pairs of correlated metrics including {M8 *(IFP), M9 (Im_DT)*}, {M9*(Im_DT)*, M10*(Sml_Cls)*} and *{*M6 *(Inc_Prp-Vlu),* M7 *(IF)}*. Given the nature of these metrics and based on their definitions presented in Section 3.3, it is not expected that the values of these metrics would be related to each other. For example, invalid usage of properties (M8) and improper data types (M9) are fundamentally independent. This reason can be applied to the other correlations. i.e. {M9 *(Im_DT),* M10 *(Sml_cls)*} and {M6 *(Inc_Prp-Vlu)*, M7 *(FP)*}.

The most likely reason for the observed correlation between the members of the three pairs is that the values for correlated metrics are mainly '0', and it indicates that there are not many deficiencies related to the mentioned metrics in the datasets. As a result, if the values of two metrics for different datasets are mainly equal to '0', we cannot conclude that those metrics are correlated. On the other hand, the results do not show transitivity between the correlations (e.g., correlation observed between M8 and M9, also between M9 and M10, but no correlation observed between M8 and M10). Hence, according to the values reported in Table 12, only 6% of the results show correlations, however we would like to perform future studies to make verify the correlation between these metrics. As it stands, the proposed metrics could be considered to be suitable for the evaluation of a dataset from different aspects given the limited amount of correlation between the metrics.

IX. CONCLUSION AND FUTURE WORKS

In this paper, a set of measurement-theoretic metrics has been proposed for the assessment of LOD dataset. In the first part of our work, we have reviewed studies that have been reported on various aspects of data quality in the main subgroups of information quality models and data quality models in the context of LOD. Subsequently, we have shown how concrete valid metrics can be developed for RDF datasets by formally representing and implementing such metrics. Defining metrics in such a formal way ensures repeatability of our experiments. The proposed metrics have been validated through both theoretical and empirical evaluations, and finally the suitability of the proposed metrics has been discussed.

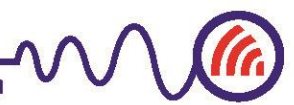

Our main findings in this paper are *I)* the identification and proper classification of quality issues that have been reported in the literature pertaining to datasets on the LOD; *ii)* the formal definition of a set of quality-driven metrics for measuring the extent of quality deficiencies in datasets prior to their publication on the LOD; and *iii)* the empirical and theoretical evaluation of these quality-driven metrics, which can be employed to shed light on distinct quality aspects of an RDF dataset.

We are currently focusing on the extension of our work in three main directions: *I)* we are working to define a set of structural metrics that could be automatically computed similarly to the quality-driven metrics proposed in this paper; *ii)* we are also considering to develop statistical models for predicting the quality dimensions of a dataset using the values of the related metrics. We have undertaken similar studies for building predictive models of quality from structural metrics in our prior research [30]; and finally *iii)* while in this paper, we have focused on the quality issues of RDF datasets which can be avoided before release, the quality issues after interlinking into the LOD remain to be further explored.

## ACKNOWLEDGMENT

The authors would like to gratefully acknowledge the support from Ryerson University's Laboratory for Systems, Software and Semantics (LS3) in particular Jelena Jovanovic for her feedback on our work, and Zoran Jeremic for his contribution in the implementation of the preliminary version of the tool.

## REFERENCES

[1] A. Zaveri, A. Rula, A. Maurino, R. Pietrobon, J. Lehmann, S. Auer, et al., "Quality Assessment for Linked Data: A Survey," Semantic Web Journal, Vol. 7(1), pp. 63-93, 2016.

[2] L. L. Pipino, Y. W. Lee, and R. Y. Wang, "Data quality assessment," Communications of the ACM, vol. 45, pp. 211-218, 2002.

[3] Y. W. Lee, D. M. Strong, B. K. Kahn, and R. Y. Wang, "AIMQ: a methodology for information quality assessment," Information & management, vol. 40, pp. 133-146, 2002.

[4] C. Batini, C. Cappiello, C. Francalanci, and A. Maurino, "Methodologies for data quality assessment and improvement," presented at the ACM Computing Surveys (CSUR), 2009.

[5] F. Naumann and C. Rolker, "Assessment methods for information quality criteria," in 5'th Conference on Information Quality Boston, Mass, USA, 2000, pp. 148-162.

[6] C. Batini and M. Scannapieca, Data quality: concepts, methodologies and techniques: Springer, 2006.

[7] B. Behkamal, M. Kahani, S. Paydar, M. Dadkhah, and E. Sekhavaty, "Publishing Persian linked data; challenges and lessons learned," in 5th International Symposium on Telecommunications (IST), 2010, pp. 732-737.

[8] Converter to RDF Available: http://www.w3.org/wiki/ConverterToRdf

[9] O. Hartig and J. Zhao, "Using Web Data Provenance for Quality Assessment," SWPM, vol. 526, 2009.

[10] C. Bizer and R. Cyganiak, "Quality-driven information filtering using the WIQA policy framework," Web Semantics: Science, Services and Agents on the World Wide Web, vol. 7, pp. 1-10, 2009.

[11] Y. Lei, A. Nikolov, V. Uren, and E. Motta, "Detecting Quality Problems in Semantic Metadata without the Presence of a Gold Standard," in EON, 2007, pp. 51-60.

[12] S. Brüggemann and F. Grüning, "Using ontologies providing domain knowledge for data quality management," in Networked Knowledge-Networked Media, ed: Springer, 2009, pp. 187-203.

[13] A. Hogan, A. Harth, A. Passant, S. Decker, and A. Polleres, "Weaving the pedantic web," in 3rd International Workshop on Linked Data on the Web (LDOW2010), Raleigh, North Carolina, 2010.

[14] C. Fürber and M. Hepp, "Using semantic web resources for data quality management," in Knowledge Engineering and Management by the Masses, ed: Springer, 2010, pp. 211-225.

[15] URI Debugger. Available: http://linkeddata.informatik.hu-berlin.de/uridbg

[16] Vapour online validator. Available: http://validator.linkeddata.org/vapour

[17] Jena Eyeball: Command line validator. Available: http://jena.sourceforge.net/Eyeball

[18] VRP: Command line validator. Available: http://139.91.183.30:9090/RDF/VRP

[19] J. Sequeda. Available: http://semanticweb.com/introduction-to-open-world-assumption-vs-closed-world-assumption_b33688

[20] ISO, "ISO/IEC 25012- Software engineering - Software product Quality Requirements and Evaluation (SQuaRE)," in Data quality model, ed, 2008.

[21] Y. Wand and R. Y. Wang, "Anchoring data quality dimensions in ontological foundations," Communications of the ACM, vol. 39, pp. 86-95, 1996.

[22] F. Naumann, U. Leser, and J. C. Freytag, Quality-driven integration of heterogeneous information systems: Citeseer, 1999.

[23] C. Batini, D. Barone, M. Mastrella, A. Maurino, and C. Ruffini, "A Framework And A Methodology For Data Quality Assessment And Monitoring," in ICIQ, 2007, pp. 333-346.

[24] V. Peralta, "Data freshness and data accuracy: A state of the art," Instituto de Computacion, Facultad de Ingenieria, Universidad de la Republica2006.

[25] J. Ashraf, O. K. Hussain, and F. K. Hussain, "A framework for measuring ontology usage on the web," The Computer Journal, 2012.

[26] M. J. Eppler and D. Wittig, "Conceptualizing information quality: A Review of Information Quality Frameworks from the Last Ten Years," in 5th International Conference on Information Quality, Boston, MA, USA, 2000, pp. 83-96.

[27] Networked Ontology (NeOn) project. Available: http://www.neon-project.org

[28] The code of metrics calculation tool Available: https://bitbucket.org/behkamal/new-metrics-codes/src

[29] J. Ashraf, O. K. Hussain, and F. K. Hussain, "Empirical analysis of domain ontology usage on the Web: eCommerce domain in focus," Concurrency and Computation: Practice and Experience, 2013.

[30] E. Bagheri and D. Gasevic, "Assessing the maintainability of software product line feature models using structural metrics," Software Quality Journal, vol. 19, pp. 579-612, 2011.

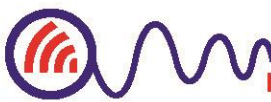

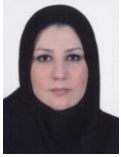
**Behshid Behkamal** is an assistant professor of Software Engineering at Ferdowsi University of Mashhad, Iran. Her research interests are the Semantic Web, Software Engineering and Data Analytics.

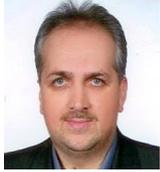
**Mohsen Kahani** is a professor of software engineering and the director of Web Technology Laboratory (WTlab) at Ferdowsi University of Mashhad, Iran. His research interests include the Semantic Web, Natural Language Processing and Software Engineering.

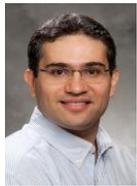
**Ebrahim Bagheri** is a Canada Research Chair in Software and Semantic Computing, Associate Professor and the Director for the Laboratory for Systems, Software and Semantics (LS3) at Ryerson University. He has expertise in the Semantic Web, Social Media/Network Analytics, and Software Engineering.

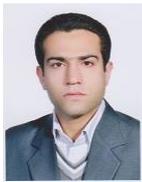
**Majid Sazvar** received the MSc degree from the Ferdowsi University of Mashhad, Iran. He received the BSc degree in computer engineering from Shamsipour Institute of Technology, Iran. His research interests include algorithm, semantic web, and bioinformatics.

IJICTR
This Page intentionally left blank.